\documentstyle[12pt,epsf]{article}
\begin{document}
\begin{titlepage}
\rightline{RU95-5-B}
\vskip 4 truecm
\centerline{ON THE IMPORTANCE OF MEASURING $\rho$ at $\sqrt{s}>1
TeV$ \footnote{To be published in the proceedings of the ``VIth
Blois Workshop, {\it Frontiers in Strong Interactions}," June 1995}}
\vspace{24pt}
\baselineskip 12pt
\centerline{N.N. Khuri}
\vspace{0.4ex}
\centerline{\it Department of Physics}
\vspace{0.4ex}
\centerline{\it The Rockefeller University, New York, New York 10021}
\vspace{3mm}
\vspace{2.0in}
\begin{abstract}
We show that, even for a soft collision like forward elastic
scattering, the
phase of the amplitude is extremely sensitive to a breakdown of strict
causality in local $QFT$.  This is especially the case when the breakdown
is manifested by a failure of polynomial boundedness which leads to
amplitudes that in some complex direction have order 1 exponential
growth in $\mid\sqrt{s}\mid$.
\end{abstract}
\end{titlepage}
\newpage
\baselineskip 18 pt

\hspace{.25in} At present the best experimental limit on the
existence of a
``fundamental length", signifying a breakdown in the local nature
of quantum
field theory (QFT), comes from (g-2) calculations and experiments in QED.
If there is a fundamental length, $R$, then from the results on the
muon (g-2)
we have

\begin{equation}
\alpha(m^2_\mu R^2)\leq 10^{-8}.
\end{equation}
This leads us to the following estimate
\begin{equation}
R^{-1}\geq O(100\;GeV).
\end{equation}
With model dependent arguments one can accommodate an $R$ such
that  $R^{-1}\approx 300-500\;GeV$, but not much better.  However, the
following assertion can be safely made:  {\it Today, we have no
experimental
evidence that can rule out the existence of a ``fundamental length," $R$,
such that $(R^{-1})>1\;TeV$}.$^{1}$

One should add that we have already reached the end of the road as
far as learning
more from QED and (g-2) regarding this issue.  At the level of
$R^{-1}\approx (1/2) TeV$
both electroweak and hadronic contributions to (g-2) become important.

The forward dispersion relations for $pp$ and ${\bar{p}p}$ scattering
represent one of the few general rigorous consequences of local quantum
field theory.  One tests their validity by measuring $\rho\equiv
ReF/Im F$,
and comparing the results with the calculated $\rho$ obtained from
the dispersion
relations with $\sigma_{tot}$ as an input.  These tests have been
carried out
at increasing  energies for both $pp$, and ${\bar{p}p}$.  The
highest energy,
${\bar{p}p}$ only, is $\sqrt{s}= 540 GeV$.  The agreement between
theory and
experiment is good, as can be seen from figure 1.

A measurement of $\rho$ at $\sqrt{s}> 1 TeV$ will explore a short
distance
domain about which we have little previous knowledge.  This is in
contrast to
the preceeding measurements which dealt with length scales that
had already
been pre-explored by QED.

However, there is a problem.  The particle physics folklore
includes statements
to the effect that short distance structure shows up first, and in
more easily
detectable ways, in \underline{hard collisions}, and not in
\underline{soft collisions}
such as forward elastic scattering.

Unfortunately, when statements like the above are repeated for 35 years,
people get to be rather sloppy in using them.  In this brief talk I
will clarify this
problem and show that in a domain where the dispersion relations no
longer hold,
 $\rho$ is extremely sensitive to the existence of a fundamental
length, $R$, and
changes drastically even when $\sqrt{s}R\approx 0.1$.

The dilemma between hard and soft collisions can be clarified as
follows.  Given
some kind of structure at short distance, $R$, we can make the
following correct
statement:

{\it  As we approach the energy region such that ${\sqrt{s} }R$ is
non-negligible,
then hard cross-sections will in general show a more detectable and
larger change
than soft cross-sections}.

This statement, though not extremely precise, is in general qualitatively
correct.  However, if in the same statement we replace
\underline{cross-sections}
by \underline{amplitudes} then the statement is false, as we shall
demonstrate by
counterexample.  Of course when local QFT is valid, then for the
forward scattering
amplitude, $F(s)$, if $\sigma_{tot} (s)$ does not change much in a
certain region
so will $ReF$ and $\rho$ because one can use the dispersion
relation to get
$\rho$  from $\sigma_{tot}$.  But, in a world in which the
dispersion relations or
QFT fail, $\rho$ as we shall show below can be extremely sensitive
to a breakdown
in QFT.

The main purpose of this talk is to give a ``counterexample" to the
statement about
soft amplitudes being insensitive to a short distance breakdown in
QFT.  Indeed,
I will show that when traditional QFT is changed by a breakdown in
polynomial
boundedness, which in many models is the main feature of the
existence of a
fundamental length, then the phase, $\rho$, could change by more
than a factor of 2
even for energies as low as $\sqrt{s} R\approx 0.1$.  This happens
without any
significant change in $\sigma_{tot}$ or $Im F$.

Before we present our ``counterexample", we summarize briefly the
well established
properties of $F(s)$ in local QFT.   We have the following:  i)
$F(s)$ is analytic
in the doubly cut s-plane;  ii) $F(s)$ is crossing symmetric,
relating $pp$ for
$s> 0$ to ${\bar{p}p}$ for $s<0$; iii) $F$ satisfies the optical theorem,
 $Im F(s)= k \sqrt{s}\sigma_{tot}> 0, s>0$;
iv)  $F(s)$ is polynomially bounded in all directions in the cut
complex s-plane,
\begin{equation}
|F(s)|\leq C|s|^N,\;\;\; |s|\rightarrow\infty.
\end{equation}

We have argued in ref. 2, that the most likely property to change
is iv), i.e.
equation (3).  Properties ii) and iii) are on a solid footing.
While even if property i),
analyticity, fails due to some new complex singularities on the
physical sheet,
these will not lead to a strong signal except for $s\rightarrow$
(new singularities).

The breakdown in iv) replaces equ. (3) by an exponential bound:
\begin{equation}
|F(s)|\leq C e^{|p_{c.m.}|R}\leq Ce^{|\frac{\sqrt{s}}{2}R| },
\;\;\; |s|\rightarrow\infty
\end{equation}

In fact we claim more.  Namely, that along some complex s direction,
$|F|$ grows exponentially like  $exp(|\frac{\sqrt{s}}{2}R |)$.  Note that
a behavior of the type $exp(|\frac{\sqrt{s}}{2}R |)^\alpha$, with
$\alpha<1$, is
excluded due to some technical mathematical arguments.  Hence one goes
from $|s|^N$ behavior to a first order exponential in
$(\frac{\sqrt{s}}{2} R)$.

There are several examples which exhibit the behavior  (4).   These are:

a.) Non-local Potential Scattering

Here one replaces the interaction term in the Schrodinger equation, by
a non-local one, i.e.,
\begin{equation}
V(x)\psi(\vec{x})\rightarrow \int d^3y V(|\vec{x}-\vec{y}|)\psi(\vec{y})
\end{equation}
where
\begin{equation}
V(|\vec{x}-\vec{y}|)\equiv 0,\;\; {\rm{for}} \;\; |\vec{x}-\vec{y}|>R.
\end{equation}

b.) Some Nonlocal Field Theories.

c.) String-Theory$^3$ ($R^{-1} \approx M_{Planck})$

Once we have exponential behavior in $\sqrt{s}$  we have a natural way to
\underline{define} a ``fundamental length".  Noting that
$(\sqrt{s}/2)\cong$ momentum
for large $s$, then the $R$ we use in the bound to make a dimensionless
exponent is by definition our fundamental length.  We take the
smallest $R$
such that for some complex sequence ${s_j}, |s_j|\rightarrow\infty$, as
$j\rightarrow\infty$, $F$ actually grows like
$exp{\mid\frac{\sqrt{s}}{2} R\mid}$.

Our main counterexample is a representative of what happens in
non-local potential
scattering as defined by equ. (5) and (6).  It also mimics the
behavior of some
non-local field theories, although these admittedly have serious
problems.

In non-local potential scattering one can easily prove the analyticity of
$F(s), s=k^2$, but $F$ is not polynomially bounded as
$|s|\rightarrow\infty$.
However, one can prove that $|F(s).(exp i k R)|< Const.$ as
$|s|\rightarrow\infty$ in all
directions .  The physical sheet here corresponds to $Imk\geq 0$,
and $R$ is defined
in equ.(6).

These facts lead us to the following ansatz:
\begin{equation}
F_t(s)\equiv F_f(s)e^{-i \frac{\sqrt{s}R}{2}},
\end{equation}
where $F_t$ is the true amplitude, and $F_f$ is a ``false"
amplitude defined
by (7).  The true amplitude satisfies the optical theorem,
$Im F_t\equiv k \sqrt{s} \sigma_{tot} > 0$,  but $Im F_f$ is not
necessarily
positive.  However, by definition it is $F_f(s)$ that is
polynomially bounded.
Also $F_f(s)$ satisfies the dispersion relations albeit with a
non-positive
$Im F_f$.  Finally,  $F_t(s)$ has no dispersion relation because of
its exponential
growth when $\sqrt{s}\rightarrow +  i\lambda$, and
$\lambda\rightarrow\infty$.

At first the ansatz(7) looks like a tautology, or at best a
definition of $F_f$.Nevertheless,  because of the special properties
of $\rho$ for
$0.1 TeV < \sqrt{s} < 0.5 TeV$ we can still learn much from this ansatz.

We consider two energy regions.  A ``low" energy region $(\sqrt{s}
R)< 0.01$,
and a ``transitional" region $0.01<(\sqrt{s} R)< 0.4$.  We also
note the fact that
for $\sqrt{s} > 0.1 TeV$, $\rho$ is small and decreases
logarithmically.  At
$\sqrt{s} = 540 GeV$, we have $\rho({\bar{p}}p) = 0.13$.  Given the
fact that
cosmic ray data tell us that $\sigma_{tot}$ will continue to increase,
$\rho_{DR}$, the dispersion relation fit shown by the curve in
figure 1,  will continue to
decrease slowly as $\sqrt{s}$ increases.

We consider the two energy regions separately.

A. Low Energy Region, $(\sqrt{s} R< 0.01)$

In this region, we have
\begin{equation}
F_t\cong F_f, \nonumber\\
\;\;\;\;\rho_t\cong \rho_f
\end{equation}

There is  no observable difference between the polynomially bounded
amplitude and the exponentially bounded one.

B. Transitional Region

Here $\sqrt{s} R$ is small but not negligible,
\begin{equation}
0.01 <(\frac{\sqrt{s}R}{2}) < 0.2.
\end{equation}

Note that $\sqrt{s}$ is still well below $(R^{-1})$.

{}From the ansatz(7), we obtain
\begin{equation}
ImF_f(s)=k\sqrt{s} \sigma_{tot}(s)[cos \frac{\sqrt{s}}{2}R+\rho_t(s)
sin \frac{\sqrt{s}}{2}R],
\end{equation}

At this stage it is important to remark that the second term in the
bracket in
(10) is doubly small,  $\rho_t$ is small and
$(sin\frac{\sqrt{s}R}{2}){\
\lower-1.2pt\vbox{\hbox{\rlap{$<$}\lower5pt\vbox{\hbox{$\sim$}}}}\ }
0.2$.  We obtain
\begin{equation}
Im F_f(s)=k\sqrt{s}\sigma_{tot}[1+
\rho_t(\frac{\sqrt{s}}{2}R)-\frac{1}{2}
{(\frac{\sqrt{s}}{2}R)}^2 + O({(\frac{\sqrt{s}}{2}R)}^3)].
\end{equation}

To proceed further let us assume that in the transitional region,
$|\rho_t|< 0.35$, a
bound almost 2.5 times larger than the value of $\rho$ at $\sqrt{s}
= 0.5 TeV$.
We hasten to add that this assumption is not really needed, but it
gives us a quick
way to arrive at our result.  Later we shall show how one can get
the same result without this assumption.  We now have,
\begin{equation}
Im F_f=k\sqrt{s} \sigma_{tot}[1-O(2\%)],
\end{equation}
for $0.01 < \frac{\sqrt{s}R}{2} <0.2$.  Thus in the transitional
region $ImF_f\cong Im F_t$,
and hence $ImF_f >0$ for  $\sqrt{s}R <0.5$ and the optical theorem
holds for $Im F_f$
to a  very good approximation.  But $F_f$ satisfies a dispersion
relation, and we
can therefore use the approximate derivative form of the dispersion
relations$^4$
to get
\begin{equation}
\rho_f(s)\cong \rho_{DR}(s),  \:  \:  {\rm{for}}  \: \: 0.01\leq
(\frac{\sqrt{s}R}{2})\leq0.2,
\end{equation}
where $\rho_{DR}(s)$ is the dispersion relation fit to $\rho$ as
the one shown by
the continuous curve in figure 1.

The ``true" and ``false" phases are related by
\begin{equation}
\phi_t(s)=\phi_f(s)-\frac{\sqrt{s}}{2} R.
\end{equation}
with $\phi= tan^{-1}(1/\rho)$.  After some simple algebra we get
\begin{equation}
\rho_t = \frac{\rho_f + u}{1 - \rho_fu}\;\;\;;\;\;\;  u \equiv tan
\frac{\sqrt{s}R}{2}
\end{equation}
But $\rho_f\cong\rho_{DR}$ in the transitional region.  Replacing
$\rho_f$ by
$\rho_{DR}$ in (15) and expanding in powers of
$(\frac{\sqrt{s}R}{2})$, we get
\begin{equation}
\rho_t = \rho_{DR} + \frac{\sqrt{s}R}{2}  + O(A^3),\;\;\; 0.01 <
\frac{\sqrt{s}R}{2} < 0.2.
\end{equation}
The error, $O(A^3)$, is determined by
\begin{equation}
A^3\equiv Max \{ \rho^2_{DR} (\frac{\sqrt{s}R}{2})\;\;  ; \;\;
\rho_{DR} (\frac{\sqrt{s}R}{2})^2\;\;;\;\;
(\frac{\sqrt{s}R}{2})^3\}.
\end{equation}

There are two important features of our final result (16) which
should be stressed.
First, while $\rho_{DR}$ decreases logarithmically the additional
term $(\sqrt{s} R/2)$
increases linearly with $\sqrt{s}$, and even when
$\frac{\sqrt{s}R}{2}\approx 0.1$ it
leads to a 75\% increase in $\rho$.  The  second feature is the
fact that $R$ appears
linearly in (16), unlike the correction to the muon (g-2) in equ.
(1) which was
$(\alpha m^2_\mu R^2)$, with $R^2$ appearing in the correction.
Since $R$ is
very small, this makes the QED case less sensitive to a fundamental
length.

In figure 1 we plot $\rho_t$ for $R^{-1}=20 TeV$.  The fact that
$\rho$ is sensitive is amply demonstrated.

In closing, we explain briefly how our assumption on $\rho_t< 0.35$
can be removed.
One has to divide the interval, $0.01\leq\frac{\sqrt{s}R}{2} \leq
0.2$, into $N$ small intervals, $N\approx O(20)$.  Then one carries
out the same calculation we did above step by step
starting at $(\frac{\sqrt{s}R}{2})=0.01$, and calculating $\rho_t$
at the next point.
At each step one uses the approximation $\rho^{(n+1)}_t\cong
\rho^{(n)}_t$.  The result
given by equ.(16) remains the same.

There are other examples which lead to a decrease in $\rho$ rather
than the
increase given by the ansatz (7).  For example one could have
instead of (7),
\begin{equation}
F_t = F_f e^{i\frac{\sqrt{s}R}{2}}.
\end{equation}

This leads to
\begin{equation}
\rho_t\cong \rho_{DR} -\frac{\sqrt{s}R}{2},
\end{equation}
with the result shown in figure 1 for $R^{-1} = 20 \;TeV$, and
labeled as case II.

Our ansatz is an example of a whole class where one can write
$F_t(s) \equiv E(\frac{\sqrt{s}R}{2})F_f(s)$, where  $\it{E}(z)$ is
an entire function
of order 1.  All these lead, via similar arguments, to a dramatic
change in $\rho$,
\begin{equation}
\rho_t = \rho_{DR}(s)\pm c(\frac{\sqrt{s}R}{2}) \;\;\; ; \;\;\;
0.01<\frac{\sqrt{s}R}{2} <0.2,
\end{equation}
where $c=O(1)$.

In conclusion, $\rho$ is extremely sensitive to a short distance
breakdown of
QFT, and more so if that breakdown is manifested by the failure of
polynomial
boundedness and  its replacement by exponential growth which is
order 1 in the
variable $(pR)$.  Here $p$ is the c.m.  momentum in forward
scattering, and $R$
we take, by definition, to be the ``fundamental length".

Strictly within the context of this picture, we can set the
following limits
on a breakdown in polynomial boundedness:

1.)  $UA 4/2$ gives us a lower limit for $R^{-1}$, $R^{-1} >7 TeV$.

2.)  $E-710$ if redone with smaller errors could improve this bound to
$R^{-1}> 20 TeV$; here $\sqrt{s} = 1.8 TeV$.

3.)  Agreement between $\rho_{th}$ and $\rho_{exp}$ at LHC could
get us $R^{-1}> 140 TeV$.

\section*{Acknowledgements}
This work was supported in part by the U.S. Department of
Energy under grant no. DOE91ER40651 TaskB.

\vspace{7mm}
\noindent{\large{\bf{References and Footnotes}}}
\vspace{5mm}

\noindent [1]  T. Kinoshita, {\it Quantum Electrodynamics}, T.K. Editor,
(World Scientific, 1990), pp. 471.  It should be stressed that one
should not confuse our $R$ here with the limits on the lepton
or quark form factors in QCD.  In axiomatic QFT  even the deuteron,
which certainly is  composite, has an interpolating field which is
strictly causal, i.e. two field operators at  x and y commute if
$(x-y)^2< 0$, i.e. spacelike.  This leads to dispersion relations for
forward $DD$ scattering.  If $\phi^\mu_D(x)$ does not satisfy
strict causality,
this will have little to do with whether $D$ or the electron are
composite or not.
\vspace {-.1 truecm}

\noindent [2]  N.N. Khuri, ``Proceedings of Les Recontres de Physiques
da la Vallee d'Aoste: {\it Results and Perspectives in Particle Physics"}
(M. Greco, Ed.), pp {\bf 701-708}.  Editions Frontieres, Gif-sur-Yvette,
France, 1994.

\noindent [3]   D.J. Gross and P. Mende, Nucl. Phys. {\bf B303},
407 (l988).
\vspace {-.1 truecm}

\noindent [4]  See for example, E. Leader, Phys. Rev. Lett. {\bf
59}, 1525
(1987).

\newpage

\begin{figure}
\ \vskip.5in
\centerline{\epsfbox{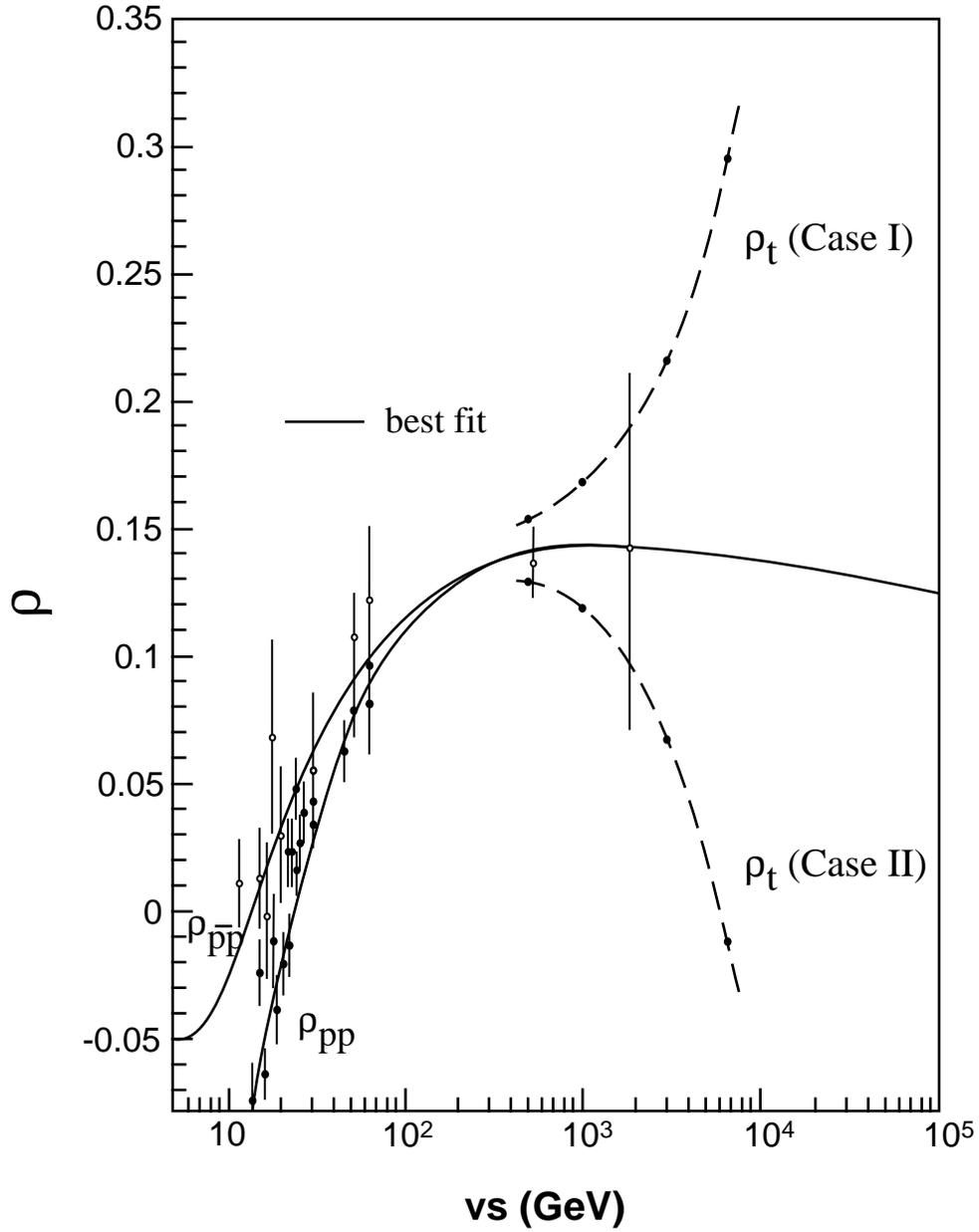}}
\bigskip
\caption{The curves give the best dispersion relation fit for
$\rho_{\overline{p}p}$ and $\rho_{pp}$.  The dashed lines represent
our calculation of $\rho_t$ for $(R^{-1})=20$ TeV, for both case I and II.}
\end{figure}

\end{document}